# The logical foundations of Gibbs' paradox in classical thermodynamics *

# V. Ihnatovych

Department of Philosophy, National Technical University of Ukraine "Kyiv Polytechnic Institute",

Kyiv, Ukraine

e-mail: V.Ihnatovych@kpi.ua

**Abstract**

The analysis of the arguments within the limits of the classical thermodynamics that lead to the Gibbs paradox was made. Features of preconditions used in the derivation of the entropy of mixing of ideal gases that caused the appearance of paradox were established. It was shown that the Gibbs paradox has not connection with the assumption of discrete differences between the parameters of different gases.

**Contents**

# 1. Introduction

Gibbs' paradox arose from theoretical reviewing of the problem of entropy change of mixing two ideal gases [1].

Considering the mixing of two different ideal gases with equal initial temperatures, pressure and the volumes, divided originally by a partition, after removal of a partition, entropy of system increases by the value of entropy of mixing of various gases $\Delta S_m$, equal $2kN\ln 2$ or $2Rn\ln 2$, where $N$ – number of molecules of each gas, k – Boltzmann' constant, $n$ – number of moles of each gas, R – universal gas constant [2–8]. This value does not depend on how much mixed gases differ. Mixing of two identical ideal gases having equal initial temperatures, pressure and the volumes, divided originally a partition, results in entropy change equal to zero. The essence of Gibbs' paradox consists in the jump of change of entropy at transition from relatively close (infinitesimal differing) by properties of gases to identical gases.

Gibbs' paradox is known for more than a century. The explanations of this paradox were offered by many physicists, among them – H. A. Lorentz, J. D. van der Waals, M. Planck, E. Fermi, A. Einstein, E. Schrödinger (see [9, 10]). There are several tens explanations of Gibbs' paradox – about fifty various solutions (explanations) which have appeared till 1986 are presented (in the monographs [9]); since then more than 80 papers devoted to Gibbs' paradox has been published (see [10]).

Though many consider that Gibbs' paradox has been explained long time ago, there is no reason to conclude that the problem is solved: there is no universal opinion what to consider as the paradox' decision; decisions of this paradox in [10] are disputable; there was no work whose author would write, that in work of other author the correct decision of Gibbs' paradox is given.

There is no common understanding in the literature what should be considered as the solution of Gibbs' paradox: one author considers, that for the decision of problem it is necessary to establish the physical foundations of jump of entropy of mixing – to find out why there is a jump [2, p.1893; 3, p.70; 11, p.24, p.201–205])], others – that the jump has to be eliminated – to find a case when entropy of mixing at transition from various gases to identical changes continuously and turns into zero without jump [6, p.52-53].

In the literature there are disagreements on the question in which theory to search for the solution of Gibbs' paradox. This paradox has been formulated and originally considered within the limits of classical thermodynamics. However subsequently the representations of statistical thermodynamics, the quantum statistics, the information theory were involved for its decision (see [5, 6, 9, 10]).

Such uncertainty is mostly caused by the authors who were engaged in search of the solution of Gibbs' paradox and have not concerned its logical aspect – searched for the decision, not taking into consideration following circumstances.

**First**. The statement about paradoxical jump of entropy of mixing is obtained not on the basis of processing of empirical data, but theoretically, by reasoning.

From logic it is known: if there are no errors in reasoning the true conclusions logically follow from premises. Accordingly, if Gibbs' paradox does not appear as a logic error in reasoning – then, regardless of existence of physical foundations of paradoxical behaviour of entropy of mixing – there should exist logical foundations of the conclusions about paradoxical behaviour of entropy of mixing, i.e. premises from which with necessity follow the conclusions about existence of jump of entropy of mixing and about the independence of size of this jump of the properties of gases. When establishing these premises, it is possible to answer many controversial questions including a question, whether the conclusion about paradoxical jump is caused by assumption of existence of discrete distinctions between parameters of mixed gases, or not.

**Second**. Gibbs' paradox is a question of ideal gases – abstract, ideal objects similar to such objects, as, for example, the triangles in geometry. Properties of ideal gases are postulated by known axioms – laws of ideal gases – or found by reasoning (calculations), – just like in geometry properties of triangles are found. Accordingly, for any of the conclusions, concerning properties of ideal gases, it is possible to establish the logical foundations.

**Third**. Gibbs' paradox is a question of peculiarities of behaviour of certain function – entropy of mixing of ideal gases which is found not by direct measurement, but by calculations under certain formulas. If to deduce the formula for entropy of mixing, then the logical foundations of the conclusion about paradoxical jump of entropy of mixing can be established quite definitely: peculiarities of behaviour of function for which the formula is known are defined by peculiarities of the formula and peculiarities of behaviour of arguments.

Thus, the purpose of the present work is to get the answer to a question: what peculiarities of which premises of known reasoning the occurrence of the conclusions about those peculiarities of mathematical behaviour of entropy of mixing of ideal gases, which the Gibbs' paradox is about, are involved?

For achievement of the specified purpose we will make a conclusion of the general formula for entropy of mixing of ideal gases and we will analyze reasoning which lead to the paradoxical conclusions. Using as initial formulas of classical thermodynamics, we will establish the logical foundations of Gibbs' paradox in classical thermodynamics.

## 2. Starting thesis and preliminary remarks

Entropy $S$ is condition function, i.e. univocal function of parameters of a condition of thermodynamic system. Entropy change $\Delta S$ at transition of system from an initial condition to a final is equal to a difference of entropy of system in final $S_{II}$ and initial $S_I$ conditions:

$$\Delta S = S_{II} - S_I. \qquad (1)$$

Hereinafter indexes $I$ designate the values concerning an initial condition of system, indexes $II$ designate the values concerning a final condition of system.

At mixing of two various ideal gases (1 and 2) an initial condition of system – the ideal gases divided by an impenetrable partition 1 and 2 which volumes are equal to $V_1$ and $V_2$; a final condition – a mixing of the gases, which volume it is equal to $V_1 + V_2$. Entropy change at mixing of various gases (entropy of mixing of various gases $\Delta S_m$) is defined by the formula:

$$\Delta S_m = S_m - S_g, \qquad (2)$$

where $S_m$ – entropy of a mixture, $S_g$ – entropy of the system consisting of subsystems, divided by impenetrable partitions.

Entropy of the system consisting of subsystems divided by impenetrable partitions, is expressed by the formula:

$$S_g = \Sigma S_j, \qquad (3)$$

where $S_j$ – entropy $j$-é subsystems.

Entropy of a mixture, according to Gibbs' theorem, is expressed by the formula:

$$S_m = \Sigma S_i, \qquad (4)$$

where $S_i$ – entropy of $i$-th ideal gas – a mixture component.

In classical thermodynamics entropy of ideal gas is expressed by following equivalent formulas:

$$S_i = n_i(c_{vi} \ln T_i + R \ln \frac{V_i}{n_i} + S_{0vi}), \tag{5}$$

$$S_i = n_i(c_{pi} \ln T_i - R \ln p_i + S_{0pi}), \tag{6}$$

$$S_i = N_i(c_{vi} \ln T_i + R \ln \frac{V_i}{N_i} + S_{0vi}), \tag{7}$$

where: $n_i$ – number of moles of *i*-th gas, $N_i$ – number of molecules of *i*-th gas, $c_{vi}$ and $c_{pi}$ – their molar thermal capacities, accordingly, at constant volume and at constant pressure, $T_i$ – thermodynamic temperature, $p_i$ – pressure of *i*-th gas, $S_{0vi}$ and $S_{0pi}$ – constants of integration which depend on the nature of gas and do not depend on $n, c, V, p, T$.

A number of parameters of *i*-th ideal gas are connected by the equation of a condition of ideal gas:

$$p_i V_i = n_i R T_i. \tag{8}$$

Let's notice, that formulas (5) – (8) are true both for pure ideal gases, and for ideal gases – components of mixtures.

Let's notice also, that pressure of a mixture of ideal gases $p_m$, according to Dalton's law, is expressed by the formula:

$$p_m = \Sigma p_i, \tag{9}$$

Where $p_i$ – partial pressure of *i*-th gas in a mixture.

## 3. Derivation and the preliminary analysis of formulas for entropy of mixing of ideal gases

**At mixing of two various ideal gases** the binary mixture is formed. If temperatures of gases before mixing are equal to $T_1$ and $T_2$ then after mixing temperature of components of a mixture is equal to temperature of a mixture $T_m$:

$$T_m = \frac{n_1 c_{v1} T_1 + n_2 c_{v2} T_2}{n_1 c_{v1} + n_2 c_{v2}}. \tag{10}$$

For this case (taking into account that the mixture volume is equal to $V_1 + V_2$) from (3) – (5) follows:

$$S_{\mathrm{I}} = S_g = n_1 c_{v1} \ln T_1 + n_2 c_{v2} \ln T_2 + n_1 R \ln \ln \frac{V_1}{n_1} + n_2 R \ln \frac{V_2}{n_2} + n_1 S_{0v1} + n_2 S_{0v2}, \quad (11)$$

$$S_{\mathrm{II}} = S_m = (n_1 c_{v1} + n_2 c_{v2}) \ln T_m + R\left(n_1 \ln \frac{V_1 + V_2}{n_1} + n_2 \ln \frac{V_1 + V_2}{n_2}\right) + n_1 S_{0v1} + n_2 S_{0v2}. \quad (12)$$

Let's transform the formula (12):

$$S_m = (n_1 c_{v1} + n_2 c_{v2}) \ln T_m + R(n_1 + n_2) \ln \frac{V_1 + V_2}{n_1 + n_2} + L_x + n_1 S_{0v1} + n_2 S_{0v2}, \quad (13)$$

where

$$L_x = -R(n_1 + n_2)\left(\frac{n_1}{n_1 + n_2} \ln \frac{n_1}{n_1 + n_2} + \frac{n_2}{n_1 + n_2} \ln \frac{n_2}{n_1 + n_2}\right). \quad (14)$$

The term $L_x$, which is function only of quantities of gases, in the literature (see for example [11]) is called a logarithmic or concentration term.

If to define:

$$x_i = \frac{n_i}{n_1 + n_2}, \quad (15)$$

where $x_i$ – molar part in a mixture of $i$-th gas,

then the formula (14) will be written as follows:

$$L_x = -R(n_1 + n_2)(x_1 \ln x_1 + x_2 \ln x_2). \quad (16)$$

From (2), (10) – (13) follows:

$$\Delta S_m = \left\{(n_1 c_{v1} + n_2 c_{v2}) \ln \frac{n_1 c_{v1} T_1 + n_2 c_{v2} T_2}{n_1 c v_1 + n_2 c_{v2}} - (n_1 c_{v1} \ln T_1 + n_2 c_{v2} \ln T_2)\right\} +$$

$$+ \left\{R(n_1 + n_2) \ln \frac{V_1 + V_2}{n_1 + n_2} - R n_1 \ln \frac{V_1}{n_1} + n_2 \ln \frac{V_2}{n_2}\right\} + L_x. \quad (17)$$

The formula (17) can be obtained in other way.

The formula for entropy change at change of a condition of $i$-th ideal gas follows from (1) and (5):

$$\Delta S_i = n_i \left(c_{vi} \ln \frac{T_{IIi}}{T_{Ii}} + R \ln \frac{V_{IIi}}{V_{Ii}}\right). \quad (18)$$

From (2) – (4) (taking into account that $S_j = S_i$), follows:

$$\Delta S_m = \Sigma \Delta S_i, \tag{19}$$

where $\Delta S_i$ – a difference of entropies of *i*-th gas before and after mixing.

From (10), (14), (18), (19) follows (17).

**At mixing of two portions of gas 1** pure gas 1 is formed. For this case, if the initial parameters of portions of gas are equal, accordingly $T_1, V_1$ and $T_2, V_2$, and quantities of gas in portions $n_1$ and $n_2$, from (1), (3), (5), (10) follows:

$$S_I = S_g = c_{v1}(n_1 \ln T_1 + n_2 \ln T_2) + n_1 R \ln \frac{V_1}{n_1} + n_2 R \ln \frac{V_2}{n_2} + (n_1 + n_2) S_{0v1}, \tag{20}$$

$$S_{II} = c_{v1}(n_1 + n_2) \ln \frac{n_1 T_1 + n_2 T_2}{n_1 + n_2} + (n_1 + n_2) R \ln \frac{V_1 + V_2}{n_1 + n_2} + (n_1 + n_2) S_{0v1}, \tag{21}$$

$$\Delta S_f = \left\{ c_{v1}(n_1 + n_2) \ln \frac{n_1 T_1 + n_2 T_2}{n_1 + n_2} - c_{v1}(n_1 \ln T_1 + n_2 \ln T_2) \right\} +$$
$$+ \left\{ R(n_1 + n_2) \ln \frac{V_1 + V_2}{n_1 + n_2} - n_1 R \ln \frac{V_1}{n_1} + n_2 R \ln \frac{V_2}{n_2} \right\}, \tag{22}$$

where $\Delta S_f$ – entropy change at mixing of identical gases.

As one would expect, in formulas (11) – (13), (17) (it taking into account (14)) and (20) – (22) there is nothing else that was not in formulas (1) – (5). As in formulas (17) and (22) there are the terms depending from $c_{vi}$ and initial temperature of gases, and $c_{vi}$ is defined by the gas nature, the value of entropy of mixing for both different and identical gases generally depends on the nature of gas. Owing to the obvious features of formulas (1), (3) – (5) formulas (17) and (22) do not contain the terms depending from $S_{0vi}$, and entropy of mixing of both various and identical gases does not depend on parameters $S_{0v1}$ and $S_{0v2}$ (like size $\Delta S_i$).

If the initial temperatures of mixed gases are equal then the values $\Delta S_m$ and $\Delta S_f$ do not depend on the nature of gas because the terms that depend on $c_{vi}$, in $\Delta S_m$ and $\Delta S_f$ turn into zero.

If to accept as one usually does when considering Gibbs' paradox, that not only initial temperatures of mixed gases are equal, but also pressures (which means that, as it follows from (8), values $(V_i / n_i)$) entropy of mixing of various gases becomes the function only of their quantities:

$$\Delta S_m = L_x. \tag{23}$$

If, besides, $n_1 = n_2 = 1$,

$$\Delta S_m = 2R \ln 2. \tag{24}$$

Under the condition of equality of initial temperatures and pressures for the case of mixing of identical gases from (22) follows:

$$\Delta S_f = 0. \tag{25}$$

Formulas (23) – (25) express briefly an essence of Gibbs' paradox: entropy of mixing of various gases **under condition of equality of their initial temperatures and pressures** depends only on quantities of gases, entropy of mixing of identical gases under the same conditions is equal to zero.

The results similar to ones that are stated by formulas (11) – (13) and (20) – (25), can be obtained if, instead of the formula (5) formula (6) is used. We will not consider them here. We will specify only, that using the formula (6) entropy of a mixture is expressed by the formula:

$$S_m = (n_1 c_{v1} + n_2 c_{v2}) \ln T_m - R(n_1 + n_2) \ln p_m + L_x + n_1 S_{0\,p1} + n_2 S_{0\,p2}, \tag{26}$$

and entropy of a component of a mixture by the formula:

$$S_i = n_i[c_{pi} \ln T_m - R \ln(x_i p_c) + S_0 p_i] = n_i(c_{pi} \ln T_m - R \ln p_m + S_{0\,pi}) - n_i R x_i. \tag{27}$$

The results given by formulas (23) – (25), can be obtained on the basis of formulas (1) – (4), (7). In this case, when using formula (7) instead of the formula (5), value $L_x$ is expressed by the formula:

$$L_x = -R(N_1 + N_2)(x_1 \ln x_1 + x_2 \ln x_2), \tag{28}$$

where

$$x_i = \frac{N_i}{N_1 + N_2}. \tag{29}$$

## 4. The analysis of behaviour of entropy of mixing at transition from mixing of different gases to mixing of identical gases

Let's assume, that transition from mixing of various gases to mixing of identical gases occurs by transition (transformation) of gas 2 in gas 1, i.e. transition from mixing of gases 1 and 2 to mixing of two portions of gas 1. Under such condition, the value of entropy of mixing changes from $\Delta S_m$, expressed by the formula (17), to $\Delta S_f$, expressed by the formula (22). We will analyze the behaviour of various terms of formula (17) at such transition and we will find out, which peculiarities of their behaviour cause occurrence of those peculiarities of behaviour of entropy of mixing about which are discussed in Gibbs' paradox.

The figure braces in formulas (17) and (22) include two terms. The first term in the formula (17) depends on thermal capacities and initial temperatures of gases. At convergence of properties of gases the peculiarities of behaviour of this terms are defined by peculiarities of behaviour of parameter $c_{v2}$ and peculiarities of the formula (17). This term may experience jump at transition to identical gases only in the case that the parameter $c_{v2}$ transforms into parameter $c_{v1}$ with jump. In case of the mixing of various gases with equal values of their molar thermal capacities this term does not change at transition from various gases to the identical. If put $c_{v2}$ equal to $c_{v1}$ in the formula (17) we will receive the first term of formula (22). It is possible to conclude, that peculiarities of behaviour of this term are not related to the paradoxical jump of entropy of mixing.

The peculiarities of the behaviour of the second term in braces of the formulas (17) and (22) are not related to the Gibbs' paradox too. This term is a function of a number of moles and initial volumes of gases, and does not depend on properties of gases and does not change at transition from mixing of various gases to mixing of identical gases.

Comparing formulas (23) and (25), and also (17) and (22), one can notice, that at transition from mixing of various gases to the mixing of identical gases logarithmic term $L_x$ transforms into zero. Jump of $L_x$ to zero causes the jump of value $\Delta S_m$ on value $L_x$ (in that specific case from $2R \ln 2$ to zero) at transition from mixing of various gases to mixing of identical gases. Jump of entropy of mixing $\Delta S_m$ does not depend on properties of mixed gases, as $L_x$ changes from value which depends only on quantities of mixed gases (according to the formula (14)) to the constant value equal to zero.

Analyzing the derivation of the formulas for entropy of mixing, one can see, that the term $L_x$ exists in the formula (13) for entropy of a mixture, however it turns into zero at transition into the pure gas entropy of which is expressed by the formula (21). There is no terms similar to $L_x$ in formulas of classical thermodynamics for internal energy, thermal capacity, pressure or temperature of a mixture of ideal gases. In classical thermodynamics the specified values at transition from various to identical gases do not undergo the paradoxical jumps. It is possible to assert therefore, that **paradoxical jump of entropy of mixture by the value $L_x$ (in that specific case from 2Rln2 to 0) at transition from mixing of various gases to mixing of identical gases is caused by behaviours of term $L_x$ which passes in the formula for entropy of mixing of various gases from the formula for entropy of a mixture of ideal gases.**

It is possible to come to the same conclusion assuming, that transition from mixing of various to mixing of identical gases occurs not by transition of gas 2 into gas 1, but by rapprochement of their properties to the subsequent transition of gas 1 and gas 2 in gas 3. In this case for various gases formulas (10) – (17) are valid, and for mixing of identical formulas (20) – (22) in which instead of parameters $c_{v1}$ and $S_{0v1}$, parameters $c_{v3}$ and $S_{0v3}$ are written down.

So that to establish the logical foundations of paradoxical jump of entropy of mixing, it is necessary to find out, what features of initial formulas cause the occurrence of term $L_x$ in the formula for entropy of a mixture, and also what premises of known reasoning cause turning of $L_x$ into zero at transition to identical gases.

## 5. The logical foundations of occurrence of logarithmic term $L_x$ in the formula for entropy of a mixture of ideal gases

Analyzing a conclusion of the formula (13) for entropy of a mixture, it is possible to observe, that presence of $L_x$ in this formula is caused, first, by term $-Rn_i \ln n_i$ in the formula (5) and, secondly, that owing to (4) terms of such kind for various gases are summarized at calculation $S_m$. (Accordingly, the occurrence $L_x$ in the formula (26) happens because the formula (27) contains the term $-Rn_i \ln x_i$ and the formula (4)). The term $L_x$ does not depend on properties of components of a mixture since term $-Rn_i \ln n_i$ ($-Rn_i \ln x_i$) do not depend on properties of gases.

Let's pay attention to a role of terms $-Rn_i \ln n_i$ that entropy of a mixture contains term $L_x$. With that circumstance, that these terms are in formulas for entropy of ideal gas and are absent in formulas for internal energy, a thermal capacity, pressure, is associated with the transition to identical gases jump on value $L_x$ is experienced by entropy of a mixture of (and entropy of mixing) ideal gases, instead of a thermal capacity or pressure. The statements are not correct that if $L_x$ «is completely based on Dalton's law» [11, c.49] or as if «physical basis of entropy term $R \ln 2$ is Dalton's law» [11, c.207]. Occurrence of term $-R n_i \ln n_i$ in the formula for entropy of pure ideal gas (5) is not connected with Dalton's law.

It is possible to say, that formula (9) similar to the formula (4) is based «completely on Dalton's law», which, however, does not contain a term similar to $L_x$ as partial pressure of ideal gas is not a logarithmic function $n_i$. The same is true for a thermal capacity and internal energy of a mixture of ideal gases.

It is possible to show a role of the term $-R n_i \ln n_i$ in occurrence of the paradoxical conclusion about jump of entropy of mixing in a following way. Some authors in the past (see for example [12]) expressed entropy of pure ideal gas by the following formula:

$$S_i = n_i (cv_i \ln T_i + R \ln V_i + S_{0vi}). \tag{30}$$

It is easy to conclude, that on the basis of the formula (30) it is impossible to get the conclusion about jump of entropy of mixing at transition from mixing of various to mixing of identical gases: using of the formula (30) identical values of entropy of mixing for various and identical gases turn up (the corresponding conclusion is available, for example, in [3]).

Certainly, if entropy of a mixture was defined not by the formula (4), but the formula of a kind (5) or (6) through corresponding parameters of a mixture (volume, pressure, a thermal capacity, etc.) the formula for entropy of a mixture would not contain term $L_x$.

Thus, occurrence in the formula for entropy of a mixture of term $L_x$ which behaviour causes jump of entropy of mixing at transition to identical gases is caused by quite obvious features of formulas listed above (3) – (5). It is one part of the logical foundations of Gibbs' paradox in classical thermodynamics.

## 6. Definition of the logical foundations of the turning of function $L_x$ into zero at transition to identical gases

According to formulas (14) and (16) (taking into account (15)), received on the basis of formulas (4) and (5), $L_x$ is function of only quantities of mixed gases $n_1$ and $n_2$. Considering transition from mixing of various to mixing of identical gases, we, like other authors, assumed, that this transition occurs at constant quantities of gases. Hence, at such transition, owing to (14) and (16), logarithmic term $L_x$ should not change. If the conclusions about jump of entropy of mixing and about turning $L_x$ into zero would be received on the basis of results of measurements, there would be a conclusion, that formulas (14) and (16), and also the theory on which basis they were received, inadequately describes behaviour of function $L_x$, and also functions $S_m$ and $\Delta S_m$ at transition from various gases to the identical. But the conclusion about jump $L_x$ to zero is received through reasoning. We should therefore conclude, that this conclusion is logically incorrect, since it is not agreed with the formulas (14) and (16) and, accordingly, with the initial formulas (4) and (5) or to specify a premises which is not contradicting with formulas (1) – (9) usage of which allows to make logically correct conclusion about turning of $L_x$ into zero at transition to identical gases.

First of all we will find out, whether it is possible in general to co-ordinate turning of $L_x$ into zero with formulas (14) and (16), accordingly, with initial formulas (1) – (9) which consequences are formulas (14) and (16).

According to (14) and (16), $L_x$ goes to 0, if sizes $n_i / (n_1 + n_2)$ (i.e. sizes $x_i$) tend to 1 or 0 (i.e. at $x_1 \to 0, x_2 \to 0$; $x_1 \to 0, x_2 \to 1$; $x_1 \to 1, x_2 \to 0$; $x_1 \to 1, x_2 \to 1$).

From (15) or (29) follows, that for $i$-th pure gas the size $x_i$ is equal 1, and for a mixture

$$\Sigma x_i = 1. \tag{31}$$

Taking into account (31) it is possible to accept, that for pure gas 1 size $x_2$ is equal to zero, and for pure gas 2 $x_1 = 0$. I.e. to define, that pure gas is a kind of a two-componential mixture, namely such two-componential mixture in which molar fraction of one component is equal to 1, and the second – to zero. Pure gas can be considered also as a special case of a multicomponent mixture (a kind of a multicomponent mixture), as such multicomponent mixture in which value of one $x_i$ is equal 1, and the others – to zero. Specific quantitative difference of pure gases and mixes is thus found out: for pure gases $x_i$ are constant, and value of one $x_i$ is equal 1, and the others – to zero, for value mixes $x_i$ can change in limits: $0 < x_i < 1$, taking into account (31).

Further,

$$x \ln x_{(x=1)} = 0, \tag{32}$$

$$\lim_{x \to 0}(x \ln x) = 0, \tag{33}$$

Whence follows:

$$\lim_{\substack{x_1 \to 1 \\ x_2 \to 0}} L_x = \lim_{\substack{x_1 \to 0 \\ x_2 \to 1}} L_x = 0 \tag{34}$$

Hence, if transition from a mixture to pure gas is carried out by moving the molar fractions of one component to 1, and the second – to zero at constant value of the sums $\Sigma n_i$ and $\Sigma x_i$ turning of $L_x$ into zero will occur according to formulas (14) and (16).

Considering (33), it is possible to accept the assumption:

$$x \ln x_{(x=0)} = 0, \tag{35}$$

with which

$$L_x(x_1 = 1, x_2 = 0) = L_x(x_1 = 0, x_2 = 1) = 0, \tag{36}$$

and the formula (25) appears to be a special case of the formula (23) at $(x_1 = 0), (x_2 = 1)$ or $(x_1 = 1), (x_2 = 0)$.

Taking into account (35), turning of $L_x$ into zero at transition to identical gases is possible to agree with the formulas (14) and (16) if specify a premise from which follows, that molar fraction of one component becomes equal 1, and the second – to zero in that case when the second component becomes identical to the first.

Such premise may be the following: **the mixture is considered at a mixture if it consists of various components; the mixture of identical gases is (it is necessary to consider) a pure gas.**

By peculiarity of this conclusion, the jump of entropy of mixing at transition from various gases to identical is caused, and also it's occurrence at the moment of transition from various gases to the identical. If properties of gases change, and quantities of gases remain constants – until distinction between mixture components remains – logarithmic term $L_x$ keeps constant value: changes of properties of gases do not influence values $x_i$ and $n_i$ and, accordingly on $L_x$. Transition from a mixture of various gases 1 and 2 to a mixture of identical gases 1 and 1, because of use of the named premise is considered transition to pure gas 1 for which $x_1 = 1, x_2 = 0$, and size $L_x$ owing to (32) and (35) turns into zero. As $x_1$ and $x_2$ turn into, accordingly, 1 and 0 by jump irrespective of with or without jump a transition from various gases to identical is performed, $L_x$ turns too into zero with jump, irrespective of the character of transition from various gases to the identical. As follows from (14), $L_x$ is a continuous value, if $x_i$ are continuous values.

Considering the logical foundations of the conclusion about jump $\Delta S_m$ at transition to identical gases, as the basis for the conclusion on a jump of parameters $x_1$ and $x_2$ it is possible to use such premise: by mixing the identical gases pure gas is formed. We used this premise as a conclusion from formulas (21) and (22). This premise is used obviously or implicitly also by other authors when deducting the formula for entropy of mixing of identical gases. For example, according to I. P. Bazarov, «for calculation of change of entropy at mixture of two portions of the same gas it is necessary to use … the direct expression for entropy of chemically homogeneous gas» [2, 3]. The basis for such statement is the premise, that at mixing of two portions of the same gas (identical gases) pure (chemically homogeneous) gas is formed. However this premise does not extend onto the case of transition from a mixture to a pure gas and cannot form the basis for the conclusion about transition of the formula (13) in the formula (21) at transition to identical gases though the above analysis shows indissoluble communication of jumps of functions $\Delta S_m$ and $S_m$. In turn, the premise «at mixing of identical gases a pure gas is formed» follows from a premise «the mixture of identical gases is pure gas». Therefore, in our opinion, it is more preferable to accept «a mixture of identical gases is pure gas» as the basis for the conclusion about jump of sizes $x_i$.

Jump $L_x$ at transition from various gases to identical gases can be compared to the jump of the sum of angles at transformation of a quadrangle to a triangle by transformation of the broken piece connecting three tops of a quadrangle, to a straight line piece. The sum of angles of a polygon is function of only numbers of its angles; the sum of angles of any polygon does not depend on degree of its difference from other polygon. Using «the quadrangle is a triangle if one of its tops lays on a straight line, connecting two other tops», or «it is necessary to consider a quadrangle as a triangle if one of its angles is equal $180°$», it is possible to achieve the conclusion about the jump of a sum (and number of) angles while during transformation the quadrangle will take the form of a triangle. Without usage of such premises it is impossible to receive the conclusion about jump of the sum of corners as there are no foundations to consider a quadrangle having the form of a triangle, a triangle, instead of «a quadrangle one angle of which is equal to $180^{\circ}$». Accordingly, without a premise «the mixture of identical gases is pure gas» (from which follows, that at transition to identical gases xi changes) it is impossible to make the conclusion about turning of $L_x$ into zero at transition to identical components if certainly, not to make an absurd assumption as if the function expressed by the formula (16), can turn into zero at values $x_i$, differing from 0 and 1, i.e. to behave not how it follows from the formula (16).

Let's consider the following. Above we interpreted transition of formulas (17) in (22) and (23) in (25) (regarding behaviour of $L_x$) as the turning of term $L_x$ into zero. There are also other interpretations in the literature.

Transition of the formula of entropy of a mixture in the formula for entropy of pure gas, by B. M. Kedrov [11, p.49, p.212], is caused by disappearance of $L_x$ in the formula for entropy of a mixture. And by I. P. Bazarov [2, p.1893; 3, p.70] – occurrence in the same formula of term $-L_x$ (it means, that the mixture of identical gases is characterized by the same, differing from 0 and 1, values of parameters $x_1$ and $x_2$, as well as a mixture of various gases of which it is formed).

B. M. Kedrov proved the disappearance of $L_x$ at transition to pure gas by disappearance of partial pressures [11, c.49, c.212]. And I. P. Bazarov proved occurrence of $-L_x$ by the necessity of usage of modified Gibbs' theorem at calculation of entropy of a mixture of identical gases [2, p.1893; 3, p.212].

B. M. Kedrov's and I. P. Bazarov's reasoning confirm, that for conclusion about jump of entropy of mixing at transition from mixing of various to mixing of identical gases, the formulas (1) – (5) are not enough, and the premises about transition from mixing of various to mixing of identical gases. On the basis of these premises it is possible to get only the conclusion about transition of the first term in the brackets of the formula (17) in similar to the term of the formula (22). For the conclusion about paradoxical jump of entropy of the mixing connected with jump of term $L_x$, one more premise (such as it has been accepted by the author of present article or such as have been accepted by the specified authors) is necessary.

But in any case the conclusion about jump of entropy of mixing follows without usage of the assumption about existence of discrete distinctions between parameters of mixed gases, hence, the assumption about possibility of continuous transition from one gas to another will not lead to elimination of conclusion about the specified jump.

In B. M. Kedrov's and I. P. Bazarov's works [2, 3, 11] many attention was given to a substantiation of why it is necessary to consider a mixture of identical gases as pure gas and is impossible to consider it as the mixture. In our opinion, it is impossible to be proved, as it is impossible to prove that it is impossible to consider a quadrangle as a quadrangle if one of its corners is equal $180°$. At the same time, if in the theory as the initial formulas (3), (5) – (7) are used, but there is no interdiction to consider pure ideal gas as a mixture of identical gases, then entropy of pure ideal gas will not be unequivocal function of a condition as when for the same condition of ideal gas it is possible to receive various values of entropy, attributing to gas various values $x_i$ which sum is equal 1.

The formulation of Gibbs' paradox which would seemingly concern features of application of Gibbs' theorem about entropy of a mixture of ideal gases is connected with violation of this interdiction.

According to (4) and (5) entropy of a mixture $xn$ and $(1-x)n$ (accordingly $n$) moles of the same gas is equal to:

$$S(xn, V, T) + S((1-x)n, V, T) = S(n, V, T) + L_x, \qquad (37)$$

where $S(n, V, T)$ – the value defined by the formula (4).

This result contradicts to the formula (4). Having found out this contradiction, J. D. van der Waals and F. Konstamm [13] have written: «... It seems, that our general principle for entropy calculation (a principle of additivity of entropy (Gibbs' theorem), expressed by the formula (5) – V.I.) is always applicable to two quantities of various gases, but is not applicable to separate portions of the same gas». This contradiction is caused by calculation of entropy of one condition of the same system is considered $x_1 = 1, x_2 = 0$, and second time – $0 < x_i < 1$, and entropy of system, being certain function, at such change of parameter changes, owing to that $(n_1 + n_2)\ln(n_1 + n_2) \ne n_1 \ln n_1 + n_2 \ln n_2$. In a similar way it is possible to come to the contradiction, calculating the sum of angles if the same geometrical figure to consider that as a triangle as a quadrangle with one angle equal $180°$.

The contradiction in definition of the entropy of pure ideal gas will not arise if accept that the mixture of identical gases is the pure gas.

On the other hand, consideration of pure gas as mixture does not influence the calculated value of its thermal capacity or internal energy just as triangle consideration as a quadrangle does not lead to the contradiction at calculation of its area or perimeter. It shows once again, that the occurrence of Gibbs' paradox is inseparably linked to that feature of the formula for entropy of ideal gas, that contains term of $n_i R \ln n_i$ type.

## 7. Conclusions

Gibbs' paradox in the form of the statement about unusual mathematical properties of entropy of mixing of ideal gases in classical thermodynamics is a logically necessary conclusion from a number of premises among which – contrary to widespread opinion – there is no premise about existence of discrete distinctions between parameters of mixed gases. Occurrence of the conclusions about the jump of entropy of mixing at transition from mixing of various to mixing of identical gases and about independence of size of this jump of the properties of mixed gases in classical thermodynamics is caused by the following statements in this theory: (I) formulas for entropy of pure ideal gas contains term $n_i R \ln n_i$; (II) entropy of a mixture of ideal gases is equal to the sum of entropy's components; (III) entropy of the system of subsystems divided by impenetrable partitions, is equal to the sum of subsystems entropies; (IV) it is necessary to consider a mixture of identical ideal gases as pure gas at which molar fraction of the basic substance is equal to 1. The physical foundations of Gibbs' paradox are the physical foundations of premises (I) – (III). The premise (IV), apparently, has no physical foundations. «Elimination» of Gibbs' paradox from classical thermodynamics is possible only by suitable replacement of any of the specified premises.